\documentstyle[12pt,epsfig,twoside,fleqn,espcrc1]{article}

\newcommand{\ben}{\begin{eqnarray}}
\newcommand{\enn}{\end{eqnarray}}
\newcommand{\be}{\begin{equation}}
\newcommand{\ee}{\end{equation}}
\newcommand{\dst}{\displaystyle\phantom{\mid}}
\newcommand{\ov}{\over\displaystyle\strut}
\def\l({\left(}
\def\r){\right)}

\title{\bf The reconstructed final state of Pb + Pb 158 AGeV reactions 
       from spectra and correlation data of NA49, NA44 and WA98 }

\author{A. Ster\thanks{E-mails: ster@rmki.kfki.hu, csorgo@sunserv.kfki.hu,
        bengt@quark.lu.se. Supported by 
        grants OTKA T024094, T026435, NWO-OTKA N25l86, OMFB-Ukraine 
	45014 and US- Hungarian MAKA 652/1998}\address{MTA 
	KFKI Research Institute for Technical Physics
	and Materials Science, \\ 
        H-1525 Budapest 114, POB 49, Hungary},
	T. Cs{\"o}rg\H o\address{MTA KFKI RMKI,
         H-1525 Budapest 114, POB 49, Hungary}
        and 
        B. L{\"o}rstad\address{Physics Department, Lund University,
        S-221 00 Lund, POB 118, Sweden}}

\begin{document}
% typeset front matter
\maketitle

\begin{abstract}
        The final state of $Pb + Pb$ reactions at CERN SPS has been
        reconstructed with the Buda-Lund hydro model,
	by performing a simultaneous fit to NA49,
        NA44 and WA98 data on particle correlations and spectra. 
\end{abstract}

\section{Introduction}

        In refs.~\cite{3d,3dqm} it was observed, for the first time,
        that the parameters of particle emitting sources
	can be determined only if {\it  
	a simultaneous analysis of the momentum distributions and
        two-particle correlation functions} is performed.
        Simultaneous fitting of particle correlations and
	spectra  were reported in 
	refs.~\cite{3dqm,3ds96,3dcf98,chap_nix,na22,na49_sc}.
        Here, we determine the reconstructed space-time picture 
	of particle emission in Pb + Pb collisions at CERN SPS 
	by fitting simultaneously the  NA44 and NA49 published data
	~\cite{na49pub,na44pub}
	on two-particle correlations and single-particle spectra
        at Pb + Pb 158 AGeV central reactions at CERN SPS.
        Preliminary data~\cite{wa98pub} from WA98 experiment 
	are also used to check the reliability and the consistency
	of the fit results.

\section{Buda-Lund hydrodynamic model}
	The Buda-Lund hydro parameterization characterizes with
	means and variances the local temperature, flow and chemical
	potential distributions of a cylindrically symmetric, finite
	hydrodynamically expanding system. 
        The four-velocity $u^\mu(x)$
         of the expanding matter is given by a scaling longitudinal Bjorken
	flow appended with a linear transverse flow, characterized by
	its mean value, $\langle u_t\rangle$.  
        A Gaussian shape of the local density
        distribution is assumed both in the transverse plane  and 
	in space-time rapidity. The changes of the inverse temperature
	are characterized with means and variances. 
	The freeze-out hypersurface is characterized by a mean
	freeze-out (proper)time $\tau_0$ and a duration parameter
	$\Delta \tau$, the variance of the freeze-out propertime distribution. 
        The following emission function $S_c(x,p)$ applies:

\ben
        S_{c}(x,p) \, d^4 x &  = & {\dst  g \ov (2 \pi)^3} \,
	\, { p^\mu d^4\Sigma_\mu(x) \ov
        \exp\l({\dst  u^{\mu}(x)p_{\mu} \ov  T(x)} -
        {\dst \mu(x) \ov  T(x)}\r) + s},
        \label{e:s} \\
	 p^\mu d^4\Sigma_\mu(x) & = &
         m_t \cosh[\eta - y]  
	H(\tau) d\tau \, \tau_0 d\eta \, dr_x \, dr_y, \\
        u^{\mu}(x) & = & \l( \cosh[\eta] \cosh[\eta_t],
        \, \sinh[\eta_t]  r_x / r_t,
        \, \sinh[\eta_t]  r_y / r_t,
        \, \sinh[\eta] \cosh[\eta_t] \r),\\
        {\dst \mu(x) \ov T(x) } & = & {\dst \mu_0 \ov T_0} -
        { \dst r_x^2 + r_y^2 \ov 2 R_G^2}
        -{ \dst (\eta - y_0)^2 \ov 2 \Delta \eta^2 }, \label{e:mu}
	\quad
        {T_0 - T_r \ov T_r} = \langle {\Delta T \over T}\rangle_r ,
	\quad	
	{T_0 - T_t \ov T_t} = \langle {\Delta T \over T}\rangle_t 
\enn
        where $\mu(x)$ is the chemical potential and $T(x)$ is the local
        temperature,
	the subscript $_c$ refers to the core of collision
	(surrounded by a halo of long lived resonances),
        the transverse mass is $m_t = \sqrt{m^2 + p_x^2 + p_y^2}$,
        the rapidity $y$ and the space-time rapidity $\eta$ are
        defined as $y = 0.5 \log\left[(E+ p_z) / (E - p_z)\right]$ and
        $\eta = 0.5 \log\left[(t+ z) / (t - z)\right]$, 
	$R_G$ stands for the transverse geometrical radius of the source,
        $ \sinh[\eta_t]  =  \langle u_t \rangle {r_t \ov R_G}$ with  
	$ r_t = \sqrt{r_x^2 + r_y^2}$ .
	The central temperature at mean freeze-out time is denoted by 
	$T_0 = T(r_x = r_y = 0; \tau = \tau_0)$.
        With the surface temperature $T_r = T (r_x = r_y = R_G,
        \tau = \tau_0)$ and the temperature after freeze-out,
        $T_t = T(r_x = r_y = 0;
        \tau = \tau_0 + \sqrt{2} \Delta\tau)$, 
        the relative transverse and temporal temperature decrease
        are introduced,
	see refs. ~\cite{3dcf98,3dqm,3d} for further details.

\subsection{Single particle spectra and two particle correlations}
        The total invariant single particle spectrum
        in $y$ rapidity and transverse mass $m_t$ is determined
	analytically using a saddle-point approximation,
\ben
		{\dst d^2 n\ov  2 \pi m_t dm_t\, dy  } & = &
		{\dst g \ov (2 \pi)^3} \,
		\overline{E} \, \overline{V} \, 
	\overline{C} 
	\, {
 	1 \ov	
        \exp\l({\dst  u^{\mu}(\overline{x})p_{\mu} 
	\ov  T(\overline{x})} -
        {\dst \mu(\overline{x}) \ov  T(\overline{x})}\r) + s} \ , 
\enn
	where $\overline{E}$ stands for the average energy,  $\overline{V}$ 
	for the average volume of the effective source of particles with a 
	given momentum $p$.  The correction factor $\overline C$  
	includes the effective intercept parameter
        $\lambda_*(y,m_t)$ of the two particle (or Bose-Einstein)
	correlation function, that controls the core ratio in the
        particle production in the core/halo picture.
        In Ref.~\cite{3d}, a Boltzmann approximation $(s = 0)$ to the above
	Invariant Momentum Distribution (IMD) as well as to the Bose-Einstein
        correlation function (BECF) was obtained in  an analytic way.
        The analytical formulas for the BECF and IMD, as were used in the
	fits, have been summarized in
        refs.~\cite{3d,3dqm,na22,3dcf98}. Bose-Einstein or Fermi-Dirac
	statistics $(s = \pm 1)$ was used in the analytic expressions 
	fitted to single particle spectra. 

\section{Fitting NA49, NA44 and WA98 Pb + Pb data}

        The kinematic parameters of the Buda-Lund model are
        fitted simultaneously to IMD and HBT radii measured
        by the CERN NA49, NA44 and WA98 experiments in central
	$Pb + Pb$ collisions at 158 AGeV.
        Core/halo correction $\propto$
        ${1 / \sqrt{\lambda_*}}$ is applied
        and the corresponding errors are propagated properly.
        Due to the these conditions unique minima are found,
	a good $\chi^2/NDF $ is obtained for all reactions,
        and the strongly coupled, normalization sensitive
        $\langle {\Delta T\over T}\rangle_t$ and $\Delta \tau$
	parameters are determined.
        On Figure 1 and 2, the fits to measured data
        are shown together with the published data. Note that the
	first 5 points of the NA44 pion spectrum were contaminated
	~\cite{na44pub} and were not included in the fit.

\begin{figure}[htb]
\begin{minipage}[t]{78mm}
%\framebox[79mm]{\rule[-26mm]{0mm}{52mm}}
\vspace*{-2.0cm}
\psfig{figure=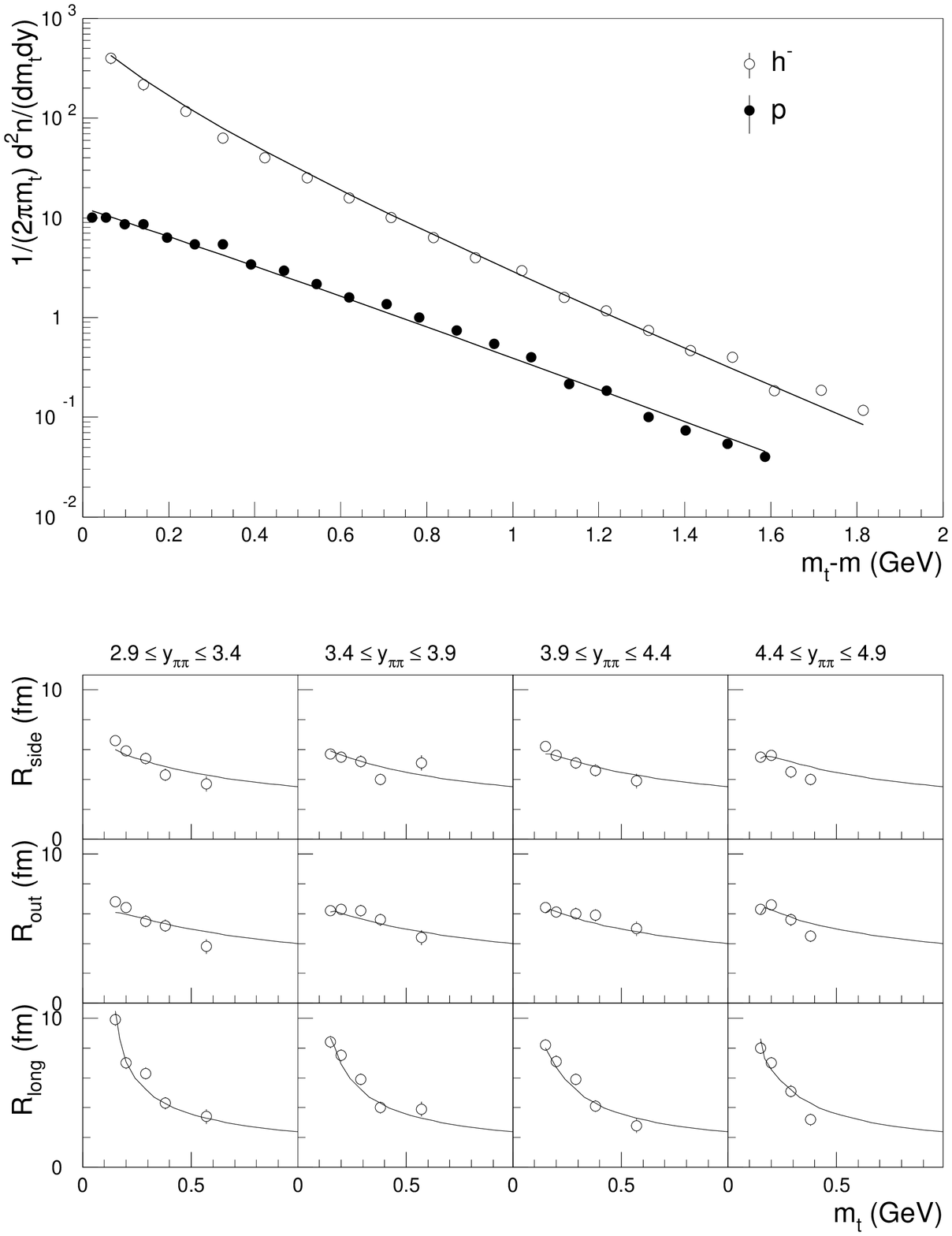,width=8.5cm,height=12.8cm}
\vspace*{-2.5cm}
\caption{Simultaneous fits to NA49 particle spectra and HBT radius parameters.}
\label{fig:na49}
\end{minipage}
\hspace{\fill}
\begin{minipage}[t]{78mm}
%\framebox[74mm]{\rule[-26mm]{0mm}{52mm}}
\vspace*{-2.0cm}
\psfig{figure=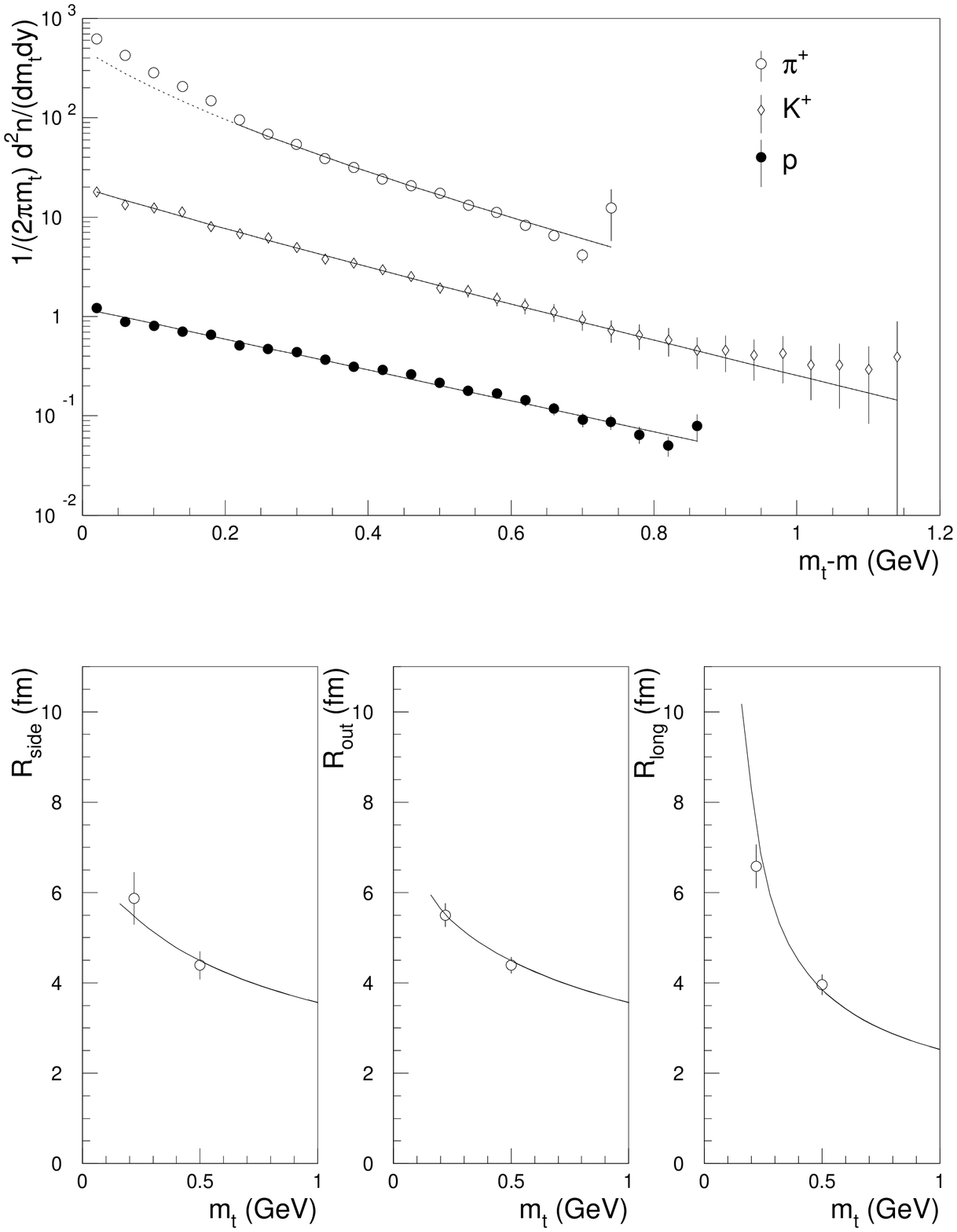,width=8.5cm,height=12.8cm}
\vspace*{-2.5cm}
\caption{Simultaneous fits to NA44 particle spectra and HBT radius parameters.}
\label{fig:na44}
\end{minipage}
\vspace*{-0.5cm}
\end{figure}

\begin{table}[hbt]
% -----------------------------------------------------
% adapted from TeX book, p. 241
\newlength{\digitwidth} \settowidth{\digitwidth}{\rm 0}
\catcode`?=\active \def?{\kern\digitwidth}
% -----------------------------------------------------
\caption{Source parameters from simultaneous fitting of 
NA49, NA44 and preliminary WA98 particle spectra and
HBT radius parameters with the Buda - Lund hydrodynamical model.}
\label{tab:results}
\begin{center}
\begin{tabular*}{\textwidth}{@{}|l@{\extracolsep{\fill}}|rl|rl|rl||rl||}
%\begin{tabular*}{@{}|l|rl|rl|rl||rl||}
\hline
                 & \multicolumn{2}{l|}{NA49} 
                 & \multicolumn{2}{l|}{NA44} 
                 & \multicolumn{2}{l||}{WA98} 
                 & \multicolumn{2}{l||}{Averaged}\\
%\cline{2-3} \cline{4-5} \cline{6-7}  \cline{8-9} 
\cline{2-9}  
		 Parameter~\,~\,~\,
                 & \multicolumn{1}{r}{Value} 
                 & \multicolumn{1}{l|}{Error} 
                 & \multicolumn{1}{r}{Value} 
                 & \multicolumn{1}{l|}{Error} 
                 & \multicolumn{1}{r}{Value} 
                 & \multicolumn{1}{l||}{Error} 
                 & \multicolumn{1}{r}{Value } 
                 & \multicolumn{1}{l||}{Error } 
		 \\
\hline
$T_0$ [MeV]      & 134  &$\pm$ 3    & 145  &$\pm$ 3   & 139  &$\pm$ 5    & 139  &$\pm$ 6   \\
$\langle u_t \rangle$ & 0.61 &$\pm$ 0.05 & 0.57 &$\pm$ 0.12 & 0.50 &$\pm$ 0.09 & 0.55 &$\pm$ 0.06 \\
$R_G$ [fm]       & 7.3  &$\pm$ 0.3  & 6.9  &$\pm$ 1.1 & 6.9  &$\pm$ 0.4  & 7.1  &$\pm$ 0.2  \\
$\tau_0$ [fm/c]  & 6.1  &$\pm$ 0.2  & 6.1  &$\pm$ 0.9 & 5.2  &$\pm$ 0.3  & 5.9  &$\pm$ 0.6  \\
$\Delta\tau$ [fm/c]  & 2.8  &$\pm$ 0.4  & 0.01 &$\pm$ 2.2  & 2.0  &$\pm$ 1.9  & 1.6  &$\pm$ 1.5  \\
$\Delta\eta$     & 2.1  &$\pm$ 0.2  & 2.4  &$\pm$ 1.6 & 1.7  &$\pm$ 0.1  & 2.1  &$\pm$ 0.4  \\
$\langle {\Delta T \over T}\rangle_r$ & 0.07 &$\pm$ 0.02 & 0.08 &$\pm$ 0.08 & 0.01 &$\pm$ 0.02 & 0.06 &$\pm$ 0.05 \\
$\langle {\Delta T \over T}\rangle_t$  & 0.16 &$\pm$ 0.05 & 0.87 &$\pm$ 0.72 & 0.74 &$\pm$ 0.08 & 0.59 &$\pm$ 0.38 \\
\hline
$\chi^2/NDF$     & \multicolumn{2}{l|}{163/98 = 1.66} 
                 & \multicolumn{2}{l|}{63/71 = 0.89} 
                 & \multicolumn{2}{l||}{115/108 = 1.06} 
                 & \multicolumn{1}{r}{1.20}
                 & \multicolumn{1}{l||}{ }\\
%$\chi^2/NDF$  & 163/98 & = 1.66 & 63/71 &= 0.89 & 115/108 & = 1.06 & 1.20 &  \\
\hline
\end{tabular*}
\end{center}
\end{table}

        Fits to preliminary data of  WA98 experiment
        provide source parameter values and errors 
	similar to those obtained by NA49 and NA44.
        The normalizations of the NA44 pion spectrum and the WA98 $h^-$spectrum
        had to be fixed manually to that of NA49.

\begin{figure}[htb]
\begin{minipage}[t]{78mm}
\vskip -2.6cm
%\framebox[34mm]{\rule[-26mm]{0mm}{52mm}}
\psfig{figure=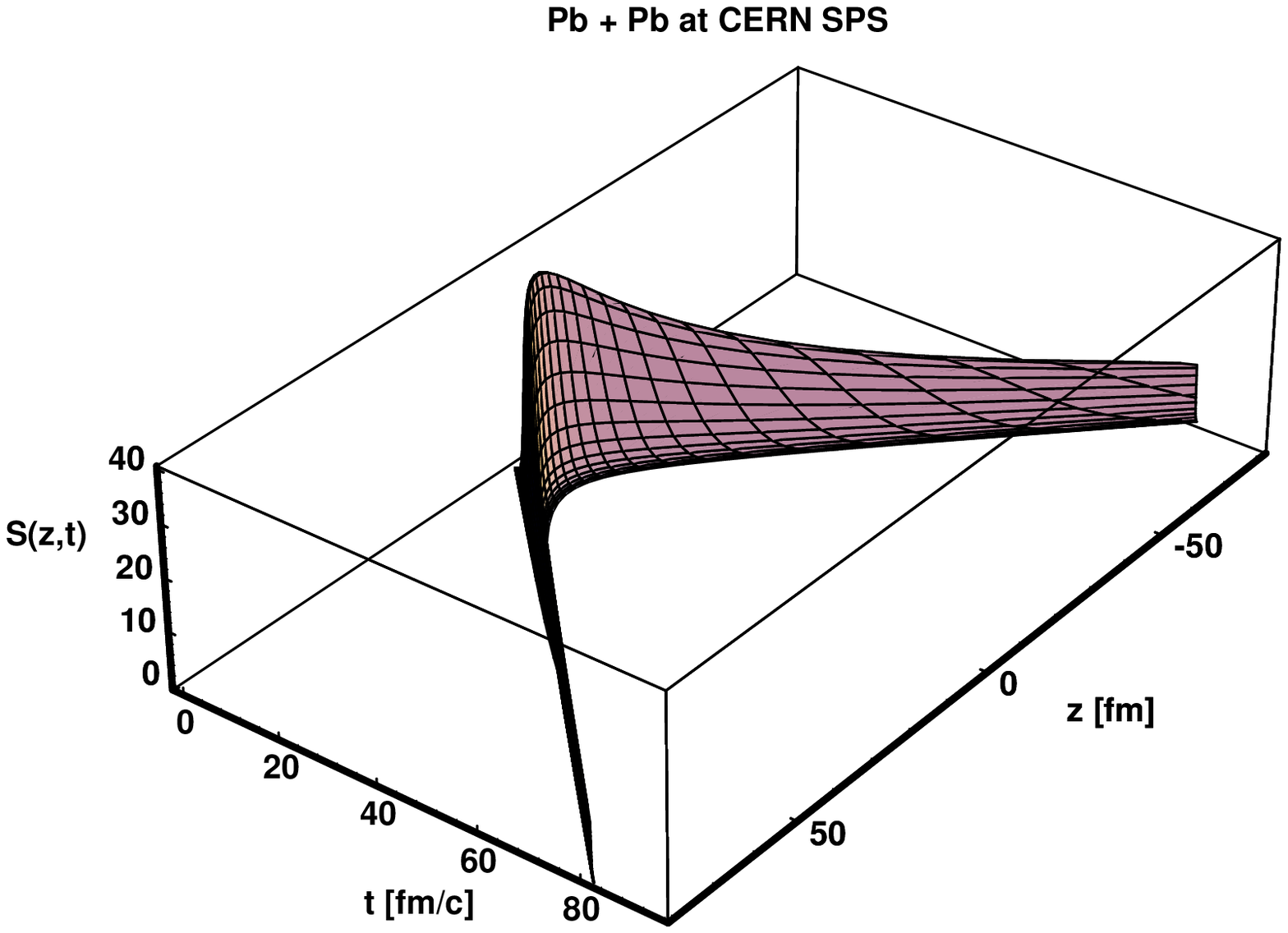,width=7.4cm,height=9.4cm}
\vskip -3.0cm
\caption{The reconstructed source function 
	$S(t,z,x=0,y=0)$ in the $(t,z)$ plane.}
\label{fig:szt}
\end{minipage}
\hspace{\fill}
\begin{minipage}[t]{78mm}
\vskip -2.6cm
%\framebox[34mm]{\rule[-26mm]{0mm}{52mm}}
\psfig{figure=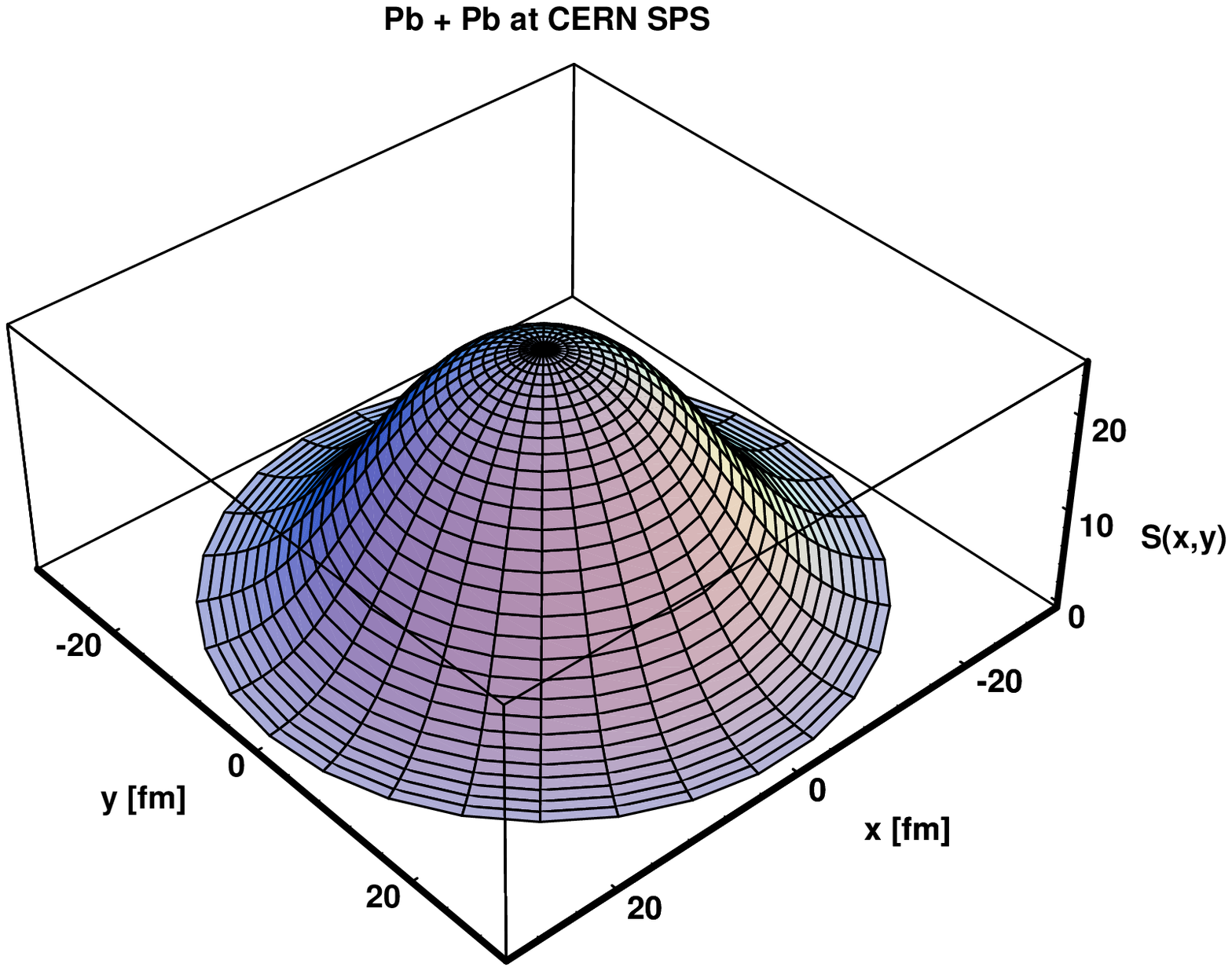,width=7.4cm,height=9.4cm}
\vskip -3.0cm
\caption{The reconstructed source function 
	$S(x,y)$ at the mean freeze-out time.} 
\label{fig:sxy}
\end{minipage}
\vskip -0.5cm
\end{figure}

	The hypothesis that
        pions, kaons and protons are emitted from the same hydrodynamical
        source is in a good agreement with all the fitted data. 
	The fit parameters are summarized in Table 1,
	shown with statistical errors, only.
	The reconsructed space-time emission function $S(x)$ 
	(which is the source function $S(x,p)$ integrated over the momentum $p$)
        is shown on Figure 3  and 4.

\section{Conclusions}
	We find that the NA49, NA44 and WA98 data on single particle
	spectra of $h^-$, identified $\pi$, $K$ and $p$ as well as 
	detailed rapidity and $m_t$ dependent HBT radius parameters
	are consistent with each other. The final state of central Pb + Pb 
	collisions at CERN SPS corresponds to a cylindrically symmetric, 
	large ($R_G = 7.1 \pm 0.2$ fm) and homogenous
	($T_0 = 139 \pm 6 $ MeV) fireball, expanding three-dimensionally
	with $\langle u_t \rangle = 0.55 \pm 0.06$. 
	A large mean freeze-out time, $\tau_0 = 5.9 \pm 0.6$ is found with 
	a short duration of emission.

\end{document}